\documentclass[AMA,STIX2COL]{WileyNJD-v2}
\usepackage{moreverb}

\newcommand\BibTeX{{\rmfamily B\kern-.05em \textsc{i\kern-.025em b}\kern-.08em
T\kern-.1667em\lower.7ex\hbox{E}\kern-.125emX}}

\usepackage{array} 
\usepackage{hyperref}
\usepackage{amsfonts}
\usepackage{amssymb} 
\usepackage{amsmath}
\usepackage{graphicx}
\usepackage{times}
\usepackage{enumerate}
\usepackage{mathrsfs}
\usepackage{verbatim} 
\usepackage{natbib}
\bibpunct{(}{)}{;}{a}{}{,} 
\usepackage{multirow}
\usepackage{color}

\articletype{Original Paper}

\received{xxx}
\revised{xxx}
\accepted{xxx}

\begin{document}

\title{Solar radius and luminosity variations induced by the internal dynamo magnetic fields}

\author[1]{F. Spada*}
\author[2]{R. Arlt}
\author[2]{M. K\"uker}
\author[3]{S. Sofia}


\authormark{SPADA \textsc{et al.}}

\address[1]{Max-Planck-Institut f\"ur Sonnensystemforschung, Justus-von-Liebig-Weg 3, 37073 G\"ottingen, Germany}
\address[2]{Leibniz-Institut f\"ur Astrophysik -- Potsdam, An der Sternwarte 16, 14482 Potsdam, Germany}
\address[3]{Department of Astronomy, Yale University, New Haven, CT 06520-8101, USA}

\corres{*Federico Spada \email{spada@mps.mpg.de}}

\def\aj{{AJ}}                   
\def\araa{{ARA\&A}}             
\def\apj{{ApJ}}                 
\def\apjl{{ApJ}}                
\def\apjs{{ApJS}}               
\def\apss{{Ap\&SS}}          
\def\aap{{A\&A}}                
\def\aapr{{A\&A~Rev.}}          
\def\aaps{{A\&AS}}              
\def\mnras{{MNRAS}}             
\def\nat{{Nature}}              
\def\ssr{{Space~Sci.~Rev.}}
\def\ao{{Appl.~Opt.}}           
\def\memsai{{Mem.~Soc.~Astron.~Italiana}} 

\renewcommand{\vec}[1]{\mbox{\boldmath$#1$}}
\newcommand{\altnote}[1]{\textsf{\textcolor{blue}{#1}}}

\abstract[Abstract]{
Although the occurrence of solar irradiance variations induced by magnetic surface features (e.g., sunspots, faculae, magnetic network) is generally accepted, the existence of intrinsic luminosity changes due to the internal magnetic fields is still controversial. 
This additional contribution is expected to be accompanied by radius variations, and to occur on timescales not limited to that of the $11$-year cycle, and thus to be potentially significant for the climate of the Earth. 
We aim to constrain theoretically the radius and luminosity variations of the Sun that are due to the effect of the variable magnetic fields in its interior associated with the dynamo cycle. 

We have extended a one-dimensional stellar evolution code to include several effects of the magnetic fields on the interior structure, such as the contributions to the hydrostatic equilibrium equation and to the energy conservation equation, the impact on the density and the equation of state, and the inhibition of convective energy transport.
We investigate different magnetic configurations, based on both observational constraints and on the output of state-of-the-art mean field dynamo models. 
We explore both step-like and simply periodic time dependences of the magnetic field peak strength. 

We find that magnetic models have decreased luminosity and increased radii with respect to their non-magnetic counterparts. 
In other words, the luminosity and radius variations are in anti-phase and in phase, respectively, with the magnetic field strength. 
For peak magnetic field strengths of the order of tens of kilogauss, luminosity variations ranging between $10^{-6}$ and $10^{-3}$ (in modulus) and radius variations between $10^{-6}$ and $10^{-5}$ are obtained. 
Modest but significant radius variations (up to $10^{-5}$ in relative terms) are obtained for magnetic fields of realistic strength and geometry, providing a potentially observable signature of the intrinsic variations. 
Establishing their existence in addition to the accepted surface effects would have very important implications for the understanding of solar-induced long-term trends on climate.
}


\maketitle

\section{Introduction}
\label{introduction}

Magnetic fields are usually not taken into account in modeling stellar structure and evolution.
Although their indirect effects are, in some cases, non-negligible (e.g., they can affect lithium depletion timescales: \citealt{Ventura_ea:1998}), their {\it direct} effects on the interior structure variables (pressure, temperature, density, etc.) and on the global stellar parameters (radius, luminosity, effective temperature) can be safely ignored, at least to the lowest order, in typical main sequence stars.
This is convenient, since the inclusion of magnetic fields in the equations describing stellar structure and evolution is very complex: nearly all equations are affected, and a self-consistent treatment would require taking into account deviations from spherical symmetry.
In some cases, however, it can be desirable, or even necessary, to include at least a first order treatment of magnetic effects in stellar models: solar-like, or ``cool", stars (i.e., of mass $M\lesssim 1\, M_\odot$) are one example.

Cool stars possess outer convection zones deep enough to sustain magnetic field generation via dynamo action; as the dynamo can feed on the rotational energy of the star, the fastest rotators are expected to host the strongest fields (see, e.g., \citealt{Ruediger_Hollerbach} for a theoretical introduction, or \citealt{Reiners:2012} for an observational review).
The direct effect of magnetic perturbations on the stellar structure may therefore be detectable, especially in young single stars, or in close binaries that are kept in a regime of fast rotation by the tidal interactions with the companion.
Indeed, discrepancies between standard, non-magnetic models and observations of the global parameters of cool stars have been known to exist for a long time \citep{Hoxie:1973,Lacy:1977}.
When sufficiently precise measurements are available (i.e., of the order of a percent or better), both single and binary stars are routinely found to have radii larger by $\approx 5$ percent, and effective temperatures cooler by $\approx 3$ percent, than the corresponding theoretical prediction from non-magnetic stellar evolution models (see, e.g., \citealt{Lopez-Morales:2007,Ribas_ea:2008,Torres_ea:2010,Boyajian_ea:2012,Feiden_Chaboyer:2012a,Spada_ea:2013}, and references therein).
Whether or not this ``radius inflation" problem is directly caused by the magnetic field of these stars remains controversial \citep{Feiden_Chaboyer:2012a}.
It is, however, intriguing that the inclusion of magnetic effects in stellar models, constructed following different theoretical approaches, has been shown to be capable of reconciling these discrepancies \citep{Chabrier_ea:2007,Feiden_Chaboyer:2012b,Feiden_Chaboyer:2013,Feiden_Chaboyer:2014,Feiden:2016}. 

Among cool stars, the Sun is a unique case: although its level of magnetic activity is modest, as expected for a slowly rotating, middle-aged main sequence star, the very precise observations available for our star can make magnetically-induced variations observable.  
In addition, understanding the role of solar variability in climate change is of great practical importance.
Since the Sun provides the vast majority of the energy input on the terrestrial climate system, even a small variation of the solar energy output on timescales of decades to centuries or longer could be a natural source of climate change \citep{Andronova_Schlesinger:2000}. 
Because solar variability on these timescales tends to be oscillatory in nature, whereas anthropogenically-induced (greenhouse) climate change is monotonic, to assess the severity of this environmental problem requires the ability to separate which process is responsible for what fraction of the change observed in the approximately $160$ years for which we have accurate climate records. 
Determining the behavior of the solar radius is significant to refine our knowledge of the magnitude and configuration of the internal magnetic field of the Sun, useful for constraining dynamo models, as well as to assess the role of solar variability on climate change, since a variation of the radius cannot occur without concomitant variations of all other global solar parameters, including the luminosity.  
If present, a variation of the sub-photospheric luminosity would contribute a fraction of the observed change of the Total Solar Irradiance (TSI; i.e., the total amount of radiation per unit surface and unit time received from the Sun by the upper atmosphere of the Earth at $1$ AU, integrated over all wavelengths), in addition to the well known changes caused by surface magnetic features \citep{Oster_ea:1982,Froehlich_Lean:2004,Froehlich:2013,Solanki_ea:2013}.

Another motivation for this work is to attempt a theoretical understanding of the solar radius variations, in order to address the many inconsistent measurements that have been reported in the literature,
both performed with ground-based facilities \citep[see the review by][]{Thuillier_ea:2005}, and operating under space-like conditions (e.g., the MDI experiment on the SOHO satellite, \citealt{Bush_ea:2010}, vs. the balloon-borne Solar Disk Sextant experiment, or SDS, \citealt{Chiu_ea:1984,Sofia_ea:1984,Sofia_ea:2013}). 
 
In this paper, we model the structural variations induced in the solar interior by the dynamo-generated magnetic fields, and study the resulting radius and luminosity variations with respect to non-magnetic models.
The magnetic effects are numerous and complex; consequently, the process of incorporating them into stellar structure models has been incremental and has focused on including only those effects deemed to be the most significant.  
Early works \citep[e.g.][]{Endal_ea:1985} have focused on investigating the response of a solar model to perturbations of the efficiency of convection and of the hydrostatic equilibrium (see also \citealt{Daeppen:1983} for an independent semi-analytical approach).
Subsequently, a self-consistent method to take into account the inhibition of convective energy transport by magnetic fields in a stellar evolution code was proposed by \citet{LS95}
(see also \citealt{Mullan_ea:2007,Mullan_ea:2012} for an alternative approach, based on the formalism of \citealt{Gough_Tayler:1966}).
At the same time, efforts have been made to improve the precision of the models, both to be sensitive to small changes, and to operate with time steps much smaller than evolutionary timescales.  

Our current approach features updated micro-physics (atmospheric boundary conditions, equation of state, opacities, element diffusion, etc.), and improvements on the formalism of \citet{LS95}.
This formalism is easily implemented in a stellar evolution code, at the cost of renouncing to model two- or three-dimensional effects.
The magnetic field is introduced in a non-magnetic, standard solar model shortly after it has reached the current solar age.
We test various internal magnetic field configurations, based on both analytical prescriptions and the output of 2D mean-field dynamo models.
Both a sudden, step-like appearance of the magnetic field, remaining constant thereafter, and a periodic evolution, with a period of $11$ years, are studied, and the evolutionary and structural effects are followed with time steps of $\approx 0.1$ yr.

This paper is organized as follows. 
In Section \ref{constraints} we describe the observational constraints that guide our modeling.  
Section \ref{method} describes the stellar code, the modifications to the stellar structure equations necessary to take the magnetic fields into account, and the magnetic configurations tested.
The results of our calculations are presented in Section \ref{sec_results}, and discussed in Section \ref{discussion}. 
Our conclusions are summarized in Section \ref{conclusions}.

\section{Observational constraints}
\label{constraints}

\subsection{Solar luminosity variations}
 
From space-based measurements, the TSI is observed to vary by approximately $0.1\%$, or $10^{-3}$, in relative terms, over a solar cycle \citep[e.g.,][]{Froehlich:2013}. 
Since the TSI measures the intensity received at the Earth, it can vary for two distinct reasons \citep{Sofia_Li:2000}: surface phenomena (i.e., sunspots, faculae, appearing and disappearing in front of the visible portion of the solar disk because of rotation, blocking or enhancing radiation in our line of sight) and the magnetic network, or intrinsic variations of the total solar luminosity.

In this work, we focus on reproducing the latter mechanism, i.e., the change of total luminosity associated with readjustments of the equilibrium structure of the Sun. 
Both sources of variation of the TSI can, of course, manifest themselves at the same time; as a consequence, the observed range of TSI variation is an upper limit for our purposes, i.e., we expect to model a luminosity variation {\it not larger} than $0.1\%$, or $(\Delta L/L)_\odot \lesssim 10^{-3}$.

\subsection{Solar radius variations}

Solar radius measurements, both from ground and from space, have a long history, which will not be discussed in detail here \citep[see, e.g.,][]{Gough:2001,Gough:2002, Thuillier_ea:2005,Thuillier_ea:2006,Djafer_ea:2008}. 

Ground based measurements of the solar radius variability often yielded results inconsistent with each other, ranging from, e.g., almost $1000$ milli-arc seconds \citep[mas;][]{Noel:2004}, to no variation at all \citep[e.g.,][]{Brown_CD:1998,Wittmann:2003}. 

Even among measurements that are not affected by the presence of the atmosphere, on the other hand, two main results exist in contradiction with each other. 
The MDI experiment onboard the SOHO satellite placed an upper limit of $23$ mas on the peak-to-peak radius variation during the $11$-year solar cycle \citep{Kuhn_ea:2004,Bush_ea:2010}; this result is usually interpreted as a non-detection.
In contrast, a significant radius variation has been detected with the Solar Disk Sextant (SDS), a stratospheric balloon--borne telescope \citep{Sofia_ea:1984}.

Although these discrepancies have not been reconciled or explained yet, a major source of concern is the absence, for most of the techniques employed, of an on-board calibration, which provides the possibility of intercomparison with measurements carried out years apart.
In this respect, a unique advantage of the SDS is the continuous onboard calibration of the instrument scale. 
Despite this advantage, the very demanding precision requirements of the measurements make the SDS data analysis a formidable challenge. 
A recent re-analysis of all the seven flights of the SDS between 1992 and 2011 with an improved reduction technique, performed by \citet{Sofia_ea:2013}, yielded a peak-to-peak solar radius variation of $\approx 200$ mas, with an uncertainty of $\pm 20$ mas over the same period (see their figure 11).
 
 In the absence of a final resolution to the controversy, we will consider the result of the SDS analysis by \citet{Sofia_ea:2013} as an upper limit on the solar radius variation: $(\Delta R/R)_\odot \lesssim 2 \cdot 10^{-4}$.

\subsection{Energy, strength, and geometry of the magnetic fields in the Sun}
\label{constraints_mf}

It is possible to place an upper limit on the energy of the magnetic fields generated by the solar dynamo within a cycle by means of a simple order of magnitude estimate \citep{Schuessler:1996, Steiner_FerrizMas:2005, Rempel:2008}.

The total magnetic flux emerging at the surface during a solar cycle is $\Phi \approx 10^{24}$--$10^{25}$ Mx \citep{Galloway_Weiss:1981}; assuming further that this flux corresponds to a thin magnetic sheet of strength $B \approx 10^5$ G located near the bottom of the solar convection zone (i.e., at $R = 0.713\, R_\odot \approx 5 \cdot 10^{10}$ cm), the magnetic energy is \citep{Rempel:2008}:
\begin{equation}
\label{total_emag}
E_{\rm mag} \approx \frac{1}{4} R\, \Phi\, B \approx 5 \cdot 10^{39}\, \mathrm{erg}.
\end{equation}
This order of magnitude estimate of $E_{\rm mag}$ will be used to assign the peak magnetic field strength in our models.

It is also instructive to compare this estimate with its counterparts for other energy reservoirs, such as the energy stored in the convective motions $E_{\rm conv}\approx 8 \cdot 10^{38}\; {\rm erg} \approx 0.16\, E_{\rm mag}$; the kinetic energy of the differential rotation $E_{\rm diff. rot.} \approx 10^{40}\; {\rm erg} \approx 2\, E_{\rm mag}$; the gravitational energy of the convection zone $E_{\rm grav} \approx 10^{47}\; {\rm erg} \approx 2\cdot 10^7\, E_{\rm mag}$; moreover, the energy associated with the observed luminosity variation during a cycle is $\Delta E_{\rm lum} \approx (\Delta L/L)_\odot \cdot L_\odot \cdot P_{\rm cycle} \approx 10^{39}\; {\rm erg}\approx 0.2\, E_{\rm mag}$ \citep{Steiner_FerrizMas:2005, Pevtsov:2012}.

A major source of uncertainty is, of course, the geometry of the field, or more specifically in our one-dimensional treatment, its radial dependence.
The best constraints on the magnetic field distribution in the solar interior available to date come from helioseismology, through the analysis of the so-called frequency splittings of the solar oscillation frequencies \citep[see, e.g.,][]{Antia_ea:2000}. 
he following recent estimates are given here as a term of comparison for the peak magnetic field strengths that will be used in our models.

Near the base of the convection zone, various authors have estimated an upper limit for the magnetic field strength ranging from $300$ kG to $1$ MG \citep{Goode_Dziembowski:1993,Antia_ea:2000,Chou_Serebryanskiy:2002,Baldner_ea:2009}. 
The results of \citet{Antia_ea:2000} are also compatible with a magnetic field strength of $20$ kG located at a depth of approximately $30000$ km below the surface, or a fractional radius of $r_0=0.960$.
At shallower depths, \citet{Baldner_ea:2009} found a best-fit of the data for a weak purely dipolar field (peak strength $=124$ G), superposed with two toroidal ``magnetic belts" located at $r_0=0.996$ and $r_0=0.999$, of intensity $1.4$ kG and $0.38$ kG, respectively.
The magnetic field strength was also found to be highly correlated with the surface activity.

Most recently, \citet{Kiefer_Roth:2018} have investigated theoretically the solar oscillation frequency shifts between the maximum and minimum of the solar activity cycle. 
From their analysis, these authors have concluded that a toroidal magnetic field layer of peak strength $\approx 40$ kG located at $r_0=0.9$ produces the best agreement of the modeled shifts with their observed counterpart.

\section{Modeling magnetic fields in the solar interior}
\label{method}

\subsection{Magnetic perturbations to the stellar structure equations}
\label{formalism}

Our modeling of the magnetic fields in the solar interior is based on the one-dimensional treatment developed over the years by the Yale solar variability group \citep[see, e.g.,][]{LS95, Li_ea:2003}.
This formalism has been implemented anew into the latest standard version of the Yale Rotational stellar Evolution Code (YREC), which includes up-to-date microphysics, such as the OPAL 2005 Equation of State \citep[EOS;][]{Rogers_Nayfonov:2002}; the OPAL Rosseland mean opacities \citep{Rogers_Iglesias:1995,Iglesias_Rogers:1996} and the \citet{Ferguson_ea:2005} low temperature opacities; helium and heavy elements diffusion  \citep{Bahcall_Loeb:1990, Thoul_ea:1994}.
The surface boundary conditions are based on the ``NextGen" batch of the \textsc{PHOENIX} model atmospheres \citep{Hauschildt_ea:1999}, available on F.~Allard's web page\footnote{\texttt{http://perso.ens-lyon.fr/france.allard/}}.
For more details on the standard input physics in YREC, see \citet{Demarque_ea:2008,Sills_ea:2000,Spada_ea:2013}.
The effects of rotation and of turbulent pressure are ignored in this work. 

In the formalism adopted here, the key physical quantity that specifies the magnetic effects in the stellar structure equations is the magnetic energy density per unit mass:
\begin{equation}
\label{chi_def} 
\chi(r,t) \equiv [B(r,t)]^2 / 8\pi \rho.
\end{equation} 
The gradient of $\chi$ is defined as: 
$$\nabla_\chi = \frac{\partial \ln \chi }{ \partial \ln P} ,$$ 
where $P$ is the total pressure, i.e., including the contribution of the magnetic pressure\footnote{\citet{LS95} introduced the magnetic pressure as $P_{\rm mag} = (\gamma-1) \rho\, \chi$, where the factor $\gamma$ (of order unity) takes into account the effect of the magnetic tension; here we adopt $\gamma=2$.} $P_{\rm mag} = \rho\, \chi$.
The total magnetic energy in the model at a given evolutionary time step is:
\begin{equation*}
U_{\rm mag}(t) = \int \frac{1}{8\pi}[B(r,t)]^2 d^3 r = \int_0^{M_\odot} \chi(r,t)\, dM_r .
\end{equation*}

The presence of magnetic fields affects all the stellar structure equations \citep[e.g., equations $10.1$--$10.4$ of][]{KWW12}. 
The effects taken into account in our formulation are summarized below.

\begin{enumerate}
\item {\it Magnetic pressure contribution}: the total pressure, $P$, replaces the non-magnetic (gas $+$ radiation) pressure $P_0$ as a stellar structure variable:
\begin{equation}
\label{modP}
P = P_0 + P_{\rm mag}.
\end{equation}
\item {\it Correction to the EOS}: the density, $\rho_0$, as obtained from a call to the standard EOS routine, is adjusted by a factor depending on the local magnetic pressure
\citep[see][]{Li_ea:2003}:
\begin{equation}
\label{modD}
\rho = \frac{\rho_0}{1 + P_{\rm mag}/P}.
\end{equation}
Equation \eqref{modD} represents a first order approximation, and we can expect it to break down in the high magnetic field regime (i.e., for $P_{\rm mag}>>P_0$). However, we have verified that all the magnetic field configurations discussed in this work are safely far from this regime, the ratio $P_{\rm mag}/P$ never exceeding $10^{-2}$.
\item
{\it Correction to the thermodynamic derivatives}: the magnetic energy density per unit mass $\chi$ appears in the EOS as an additional state variable:
\begin{equation}
\label{modEOS}
\frac{d\rho}{\rho} = \alpha \frac{dP}{P} - \delta \frac{dT}{T} - \nu \frac{d\chi}{\chi};
\end{equation}
as a consequence, the thermodynamic derivatives $\alpha$, $\delta$ are redefined, and the new quantity $\nu$ is introduced:
\begin{subequations}
\label{modEOSd}
\begin{align}
\alpha &\equiv \left( \frac{\partial \ln \rho}{\partial \ln P} \right)_{(T,\chi)}; 
\\
\delta &\equiv \left( \frac{\partial \ln \rho}{\partial \ln T} \right)_{(P,\chi)}; 
\\ 
\nu &\equiv \left( \frac{\partial \ln \rho}{\partial \ln \chi} \right)_{(P,T)};
\end{align}
\end{subequations}
the specific heat at constant pressure, $c_P$, and the adiabatic gradient, $\nabla_{\rm ad}$, are also modified \citep[see][and \ref{app_derivatives} for more details]{LS95}.
\item 
{\it Modified treatment of convection}: the standard description of convection in \textsc{YREC}, based on the mixing length theory (MLT: \citealt{BV58}; see also \citealt{Paczynski:1969}) is modified according to the method of \citet{LS95}; the magnetic field affects the local criterion for convective instability: 
\begin{equation}
\label{modG}
\nabla_{\rm rad} > \nabla_{\rm ad} - \frac{\nu}{\alpha} \nabla_{\rm ad} \nabla_\chi.
\end{equation}
The calculation of the local temperature gradient and local convective velocity are also affected (see \citealt{LS95,Li_ea:2003} for details).
\item 
{\it Variable magnetic energy contribution}: when the magnetic field is variable in time, the change to the magnetic energy should be included in the total energy budget. 
We model this effect as an additional energy source/sink $\varepsilon_{\rm mag}$ (in units of erg g$^{-1}$ s$^{-1}$) in the energy equation \citep[cf. equation 10.3 of][]{KWW12}: 
\begin{equation}
\label{modE}
\varepsilon_{\rm mag} \propto -\frac{d}{dt} U_{\rm mag}.
\end{equation}
This contribution is assumed to be equally distributed over all the shells contained within the convection zone of our models.
The term \eqref{modE} has a very modest effect on the final results, and it is only included for consistency in modeling the magnetic perturbation.
Its prescription is admittedly a crude representation of the ultimate re-coupling of the magnetic energy with the thermal energy reservoir. 
Such a process, which in reality happens through the interaction of the magnetic field with differential rotation and convective motions, cannot be described in detail in a 1D stellar model, which does not take into account these dynamical effects.
\end{enumerate}

In specifying the functional dependence of the magnetic field on the radial coordinate $r$ and the time $t$, we choose a factorable dependence for simplicity:
\begin{equation}
B(r,t) = B_0 \, f(r) \, g(t),
\end{equation}
where $B_0$ is an overall scale factor, and both $f$ and $g$ are normalized to unity.

The simplifying assumption of a factorable dependence on $r$ and $t$ is not intrinsically required by our formalism, and could be abandoned in favor of a more realistic description, for example to take into account the radial variation of the magnetic field with the phase of the cycle caused by the upward propagation of the dynamo wave.

\begin{figure}
\begin{center}
\includegraphics[width=0.5\textwidth]{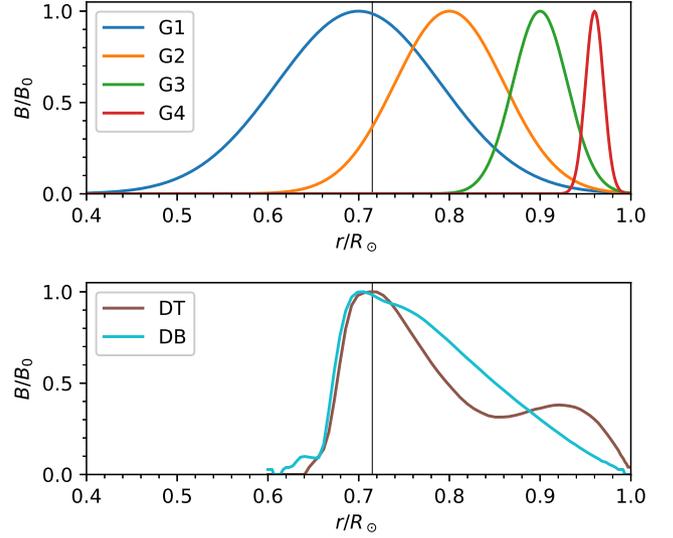}
\caption{Radial profile of the magnetic field in our models (Gaussian, G1--G4; Dynamo ``Top" and ``Bottom", DT and DB, respectively; cf. Table \ref{chiparm}). The vertical black line marks the lower boundary of the outer convection zone.}
\label{chi_radial}
\end{center}
\end{figure}

\begin{table}
\caption{Parameters of the $f(r)$ profiles for the Gaussian models (G1--G4, equation \ref{gaussian}) and the dynamo models (DT, DB, equation \ref{dynchi}).}
\begin{center}
\begin{tabular}{ccccc}
\hline
\hline
Model & $B_0$ (kG) & $r_0$ ($R_\odot$) & $\sigma_0$ & $E_{\rm mag}$ (erg) \\
\hline
G1 & $19.4$  & $0.70$ & $0.090$ & $5 \cdot 10^{39}$ \\
G2 & $20.8$  & $0.80$ & $0.060$ & $5 \cdot 10^{39}$ \\
G3 & $26.2$  & $0.90$ & $0.030$ & $5 \cdot 10^{39}$ \\
G4 & $38.9$  & $0.96$ & $0.012$ & $5 \cdot 10^{39}$ \\ 
\hline
DB & $19.5$   & N/A & N/A           & $5 \cdot 10^{39}$ \\
DT & $22.1$   & N/A & N/A           & $5 \cdot 10^{39}$ \\
\hline
\end{tabular}
\end{center}
\label{chiparm}
\end{table}

\subsection{The radial profile of the magnetic perturbation}
\label{bprofile}

To specify the radial dependence of $f$ we adopt the following two alternative approaches:
\begin{enumerate}
\item An analytical prescription, as a Gaussian function of the radial coordinate $r$.
This is controlled by two parameters: the depth of the maximum, $r_0$, and its width, $\sigma_0$ (cf. \citealt{Feiden:2016}): 
\begin{equation}
\label{gaussian}
f(r) = \exp \left[ -\frac{(r-r_0)^2}{2\sigma_0^2} \right].
\end{equation}
This simple radial dependence is ideal to explore the impact on the models of magnetic layers concentrated at different depths.

\item A numerical approach, based on the output of a two-dimensional mean-field dynamo code (described in more detail in Section \ref{dynamo}):
 \begin{equation}
\label{dynchi}
f(r) = \frac{1}{\cal N} \frac{\langle B^2\rangle}{8\pi\rho},
\end{equation}
where $\langle B^2\rangle$ is the (one-dimensional) average of $B^2$ over spherical surfaces, as obtained from the 2D dynamo model:
\begin{equation}
\langle B^2\rangle(r) = \frac{1}{4\pi} \int\limits_0^\pi B^2(r,\vartheta) \sin \vartheta d\vartheta,
\end{equation}
and $\cal N$ is a normalization factor.
\end{enumerate}

Two dynamo field configurations were calculated. 
In the ``Dynamo Bottom" (or DB) configuration, the magnetic field is maximum just below the bottom of the convection zone, and decreases monotonically with radius, whereas the ``Dynamo Top", or DT, model, features a relative maximum at shallower depth and a less steep decrease towards the surface (see Section \ref{dynamo}).

In all our models, we prescribe the overall magnetic field peak strength $B_0$ based on the assumption (cf. equation \ref{total_emag}):
\begin{equation}
\label{Emag}
\frac{B_0^2}{8\pi} \int [f(r)]^2 d^3 r \equiv E_{\rm mag} = 5\cdot 10^{39}\; {\rm erg}.
\end{equation}
Note that for both the DT and the DB configurations the dynamo code gives a peak field strength roughly in equipartition with the energy of the convective motions: $B_{\rm max} \approx 2 \, B_{\rm eq}$ (as defined in equation \ref{b_equipart}).
Moreover, the scale factor between $B_{\rm max}$ and $B_{\rm eq}$ depends on the assumptions on the intensity of the $\alpha$-effect (cf. Section \ref{dynamo}), i.e., it is model-dependent. 
For these reasons, we adopt the observationally motivated normalization in equation \eqref{Emag} for the dynamo models as well.  

The parameters of the models are summarized in Table \ref{chiparm}, and their radial profiles are plotted in Figure \ref{chi_radial}.

\subsection{Time dependence of the magnetic perturbation}

Our evolutionary calculations begin from a calibrated standard solar model (SSM), of fiducial age $t_\odot=4.57$ Gyr; the magnetic perturbation is introduced at a time $t_0\gtrsim t_\odot$ (see Section \ref{sec_results}).
We have implemented two different forms of $g(t)$:
\begin{enumerate}
\item Step-like:
\begin{equation}
\label{step}
g(t) = \left\{
\begin{array}{cc}
 0 &  , \ \ t < t_0   \\
 1  &  , \ \  t\geq t_0   \\  
\end{array}
\right. .
\end{equation}
\item Periodic with period $P=22$ yr:
\begin{equation}
g(t) = \left\{
\begin{array}{cc}
 0 &  , \ \ t < t_0   \\
 \sin \frac{2\pi t}{P}
 &  , \ \  t\geq t_0   \\  
\end{array}
\right. .
\label{periodic}
\end{equation}
\end{enumerate}


\begin{table}[b]
\caption{Dynamo model parameters. DB is a model with $\alpha$ confined to the base of the convection zone, while DT is a model with $\alpha$ located in the upper half of the convection zone.}
\begin{center}
\begin{tabular}{ccc} 
\hline
\hline
Parameter     &  DB Model &  DT Model \\
\hline
$C_\Omega$    &     $3\cdot10^4$   &   $3\cdot10^4$  \\
${\rm Rm}$    &     $300$          &   $100$         \\
$C_\alpha$    &     $3$            &   $8$           \\
$r_1$         &     $0.68$           &   $0.85$          \\
$r_2$         &     $0.75$          &   $0.97$          \\
\hline
$\eta_{\rm core}$&\multicolumn{2}{c}{0.01}           \\
$d$              &\multicolumn{2}{c}{0.02}           \\
$r_{\rm b}$      &\multicolumn{2}{c}{0.713}          \\
\hline
\end{tabular}
\end{center}
\label{dynamos}
\end{table}

\subsection{The dynamo models}
\label{dynamo}

Two different prescriptions for the radial dependence of the $\alpha$-effect were considered in our calculations. 
In the DB case the $\alpha$-effect is centered at the bottom of the convection zone, between $r_1=0.68$ and $r_2=0.75$, while the DT case has $r_1=0.85$ and $r_2=0.97$. 
Both configurations are designed to produce oscillatory solutions with equatorward migration of the azimuthal magnetic field near the bottom of the convection zone. 
All the parameters are listed in Table~\ref{dynamos}. 

\begin{figure*}
\begin{center}
\includegraphics[width=0.45\textwidth]{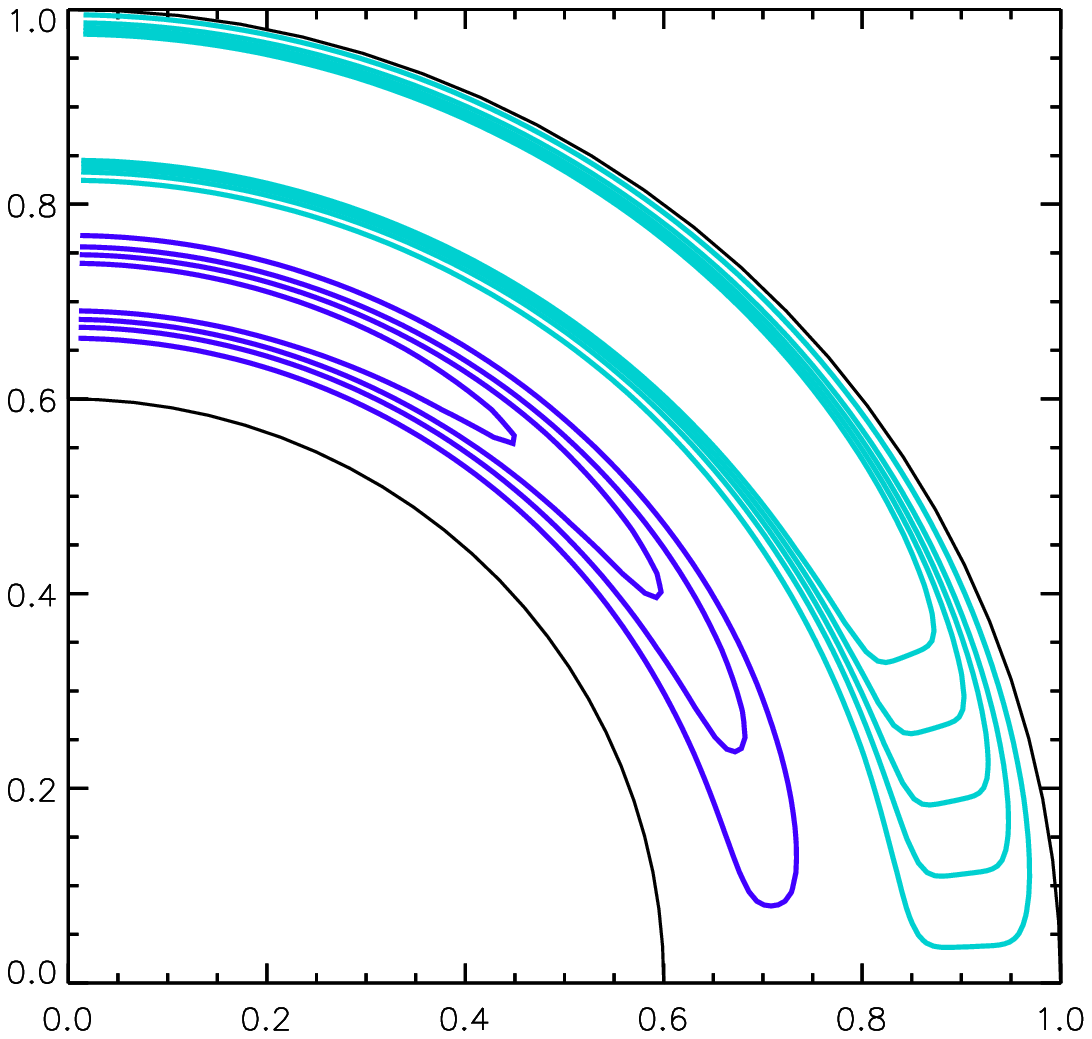}
\includegraphics[width=0.45\textwidth]{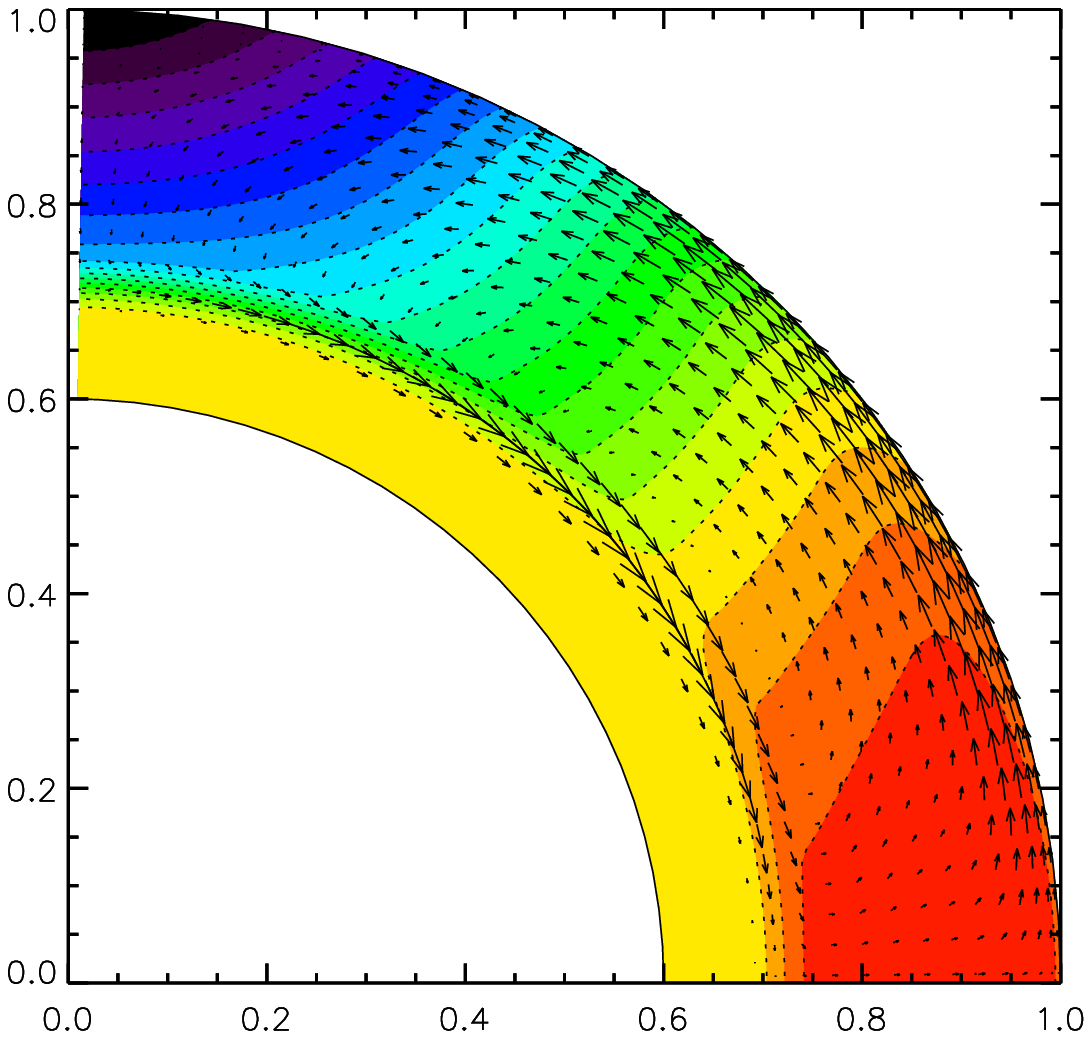}
\caption{Main sources of the magnetic amplification in our dynamo models: $\alpha$-effect and large-scale gas motions (i.e., differential rotation and meridional circulation). Left panel: contours of the $\alpha$-effect in the DB and DT models (violet and light blue, respectively). Right panel: angular velocity contours (colors) and meridional circulation (arrows), from a state-of-the-art mean-field modeling \citep{kueker2011}.}
\label{alpha_omega}
\end{center}
\end{figure*}

The amplification of the large-scale magnetic field is based on shear, and on the $\alpha$-effect, a component of the average turbulent electromotive force generated by rotating, stratified turbulence (kinetic helicity: \citealt{Parker:1955,Steenbeck_ea:1966}).
The $\alpha$-effect only produces poloidal magnetic fields -- a choice suitable for rotational velocity faster than an $\alpha$-effect of a few tens of m s$^{-1}$, usually termed ``$\alpha\Omega$-dynamo" regime \citep{Steenbeck_Krause:1966}. 
We do not attempt here to model the solar dynamo in particular; we rather compute two magnetic field distributions from two different interior profiles of the $\alpha$-effect, to get an idea of the impact on solar radius variations. 

For the $\alpha\Omega$-dynamo, the normalized mean-field induction equation reads:
\begin{eqnarray}
  \frac{\partial\vec{B}}{\partial t} & = &
  \nabla\times\Bigl[(C_\Omega u_\varphi\vec{e}_\varphi+{\rm Rm}\,\vec{u}_{\rm p})\times\vec{B} + 
  C_\alpha\psi\alpha B_\varphi \vec{e}_\varphi 
\nonumber\\
  &-& \sqrt{\eta_{\rm T}}\,\nabla\times (\sqrt{\eta_{\rm T}}\vec{B})\Bigr],
  \label{normalized}
\end{eqnarray}
where the following dimensionless parameters have been introduced: 
\begin{equation*}
C_\Omega = R_\odot^2\Omega_{\rm eq}/\eta_{\rm cz}, \ \ \ {\rm Rm}=R_\odot u_{\rm max}/\eta_{\rm cz}, \ \ \  C_\alpha = R_\odot \alpha_0/\eta_{\rm cz} ,
\end{equation*}
with $\Omega_{\rm eq}$ being the equatorial surface angular velocity, $u_{\rm max}$ the maximum meridional flow, $\alpha_0$ the amplitude of the $\alpha$-effect in the Sun, and $\eta_{\rm cz}$ the turbulent magnetic diffusivity in the convection zone. 
Note that $C_\Omega \gg C_\alpha$ for an $\alpha\Omega$-dynamo.

The first term on the right-hand side of equation \eqref{normalized} is the direct induction by the given rotational velocity $u_\varphi$ and the meridional circulation $\vec{u}_{\rm p}$; the second term is the small-scale induction effect as described by the $\alpha$-effect, and the third term comprises actually two effects: the turbulent magnetic diffusion and the diamagnetic pumping of large-scale magnetic fields against gradients in the turbulence intensity. 
The latter effect acts like a velocity, ``pumping" magnetic fields downwards in case the diffusivity (i.e. turbulence intensity) increases with radius when leaving the radiative interior and entering the convection zone \citep{kr1980,ko2012}.

The reason why the normalized magnetic diffusivity $\eta_{\rm T}$ ($=1$ in the convection zone) remains in the equation is its spatial dependence: the radial profile of $\eta_{\rm T}$ is assumed to be
\begin{equation}
  \eta_{\rm T} = \eta_{\rm core} + \frac{1-\eta_{\rm core}}{2}
  \left[1+{\rm erf}\left(\frac{r-r_{\rm b}}{d}\right)\right],
\end{equation}
where $\eta_{\rm core}=0.01$ mimics the low microscopic diffusivity in the radiative interior.

The flow profiles for the rotation, $u_\varphi(r,\vartheta)$, and the meridional circulation, $\vec{u}_{\rm p}=(u_r,u_\vartheta,0)$, are taken from a state-of-the-art mean-field modeling of the solar differential rotation \citep[$\Lambda$-effect, cf.\ ][]{kueker2011}. 
The latter technique is supported by helioseismic inversions of the solar internal rotation and observational data from the Kepler mission \citep{reinhold2013}, and we do not need to use fictional flow profiles, unlike many other solar dynamo models. 

Its spatial distribution is assumed to be: 
\begin{equation}
\label{alphaeff}
  \alpha =  \frac{1}{2}\cos\vartheta\left[1+{\rm erf}\left(\frac{r-r_1}{d}\right)\right]
  \left[1-{\rm erf}\left(\frac{r-r_2}{d}\right)\right],
\end{equation}
where $r_1$ and $r_2$ define the radial window in which the $\alpha$-effect operates. 
A smooth transition from $\alpha$-free regions occurs over the width $d$. The suppression of the $\alpha$-effect by strong magnetic fields is described by \citep[see, e.g.][]{brb1989}:
\begin{equation}
  \psi(B^2) = \frac{1}{1+\left(B/B_{\rm eq}\right)^2},
\end{equation}
where it is assumed that this suppression takes place essentially for magnetic fields larger than the equipartition field strength:
\begin{equation}
\label{b_equipart}
B_{\rm eq} = \left( 4\pi \rho v_{\rm turb}^2 \right)^{1/2}.
\end{equation}
This leads to dynamo fields of the order of $B_{\rm eq}$, but it is not a normalization of the field.

A comprehensive compilation of setups is given in Chapter~4 of \cite{charb2010}. 
For the two dynamos employed here, we use the induction equation integration from the magneto-convection scheme by \cite{hollerbach2000}.


\section{Results}
\label{sec_results}

\begin{figure*}
\begin{center}
\includegraphics[width=0.9\textwidth]{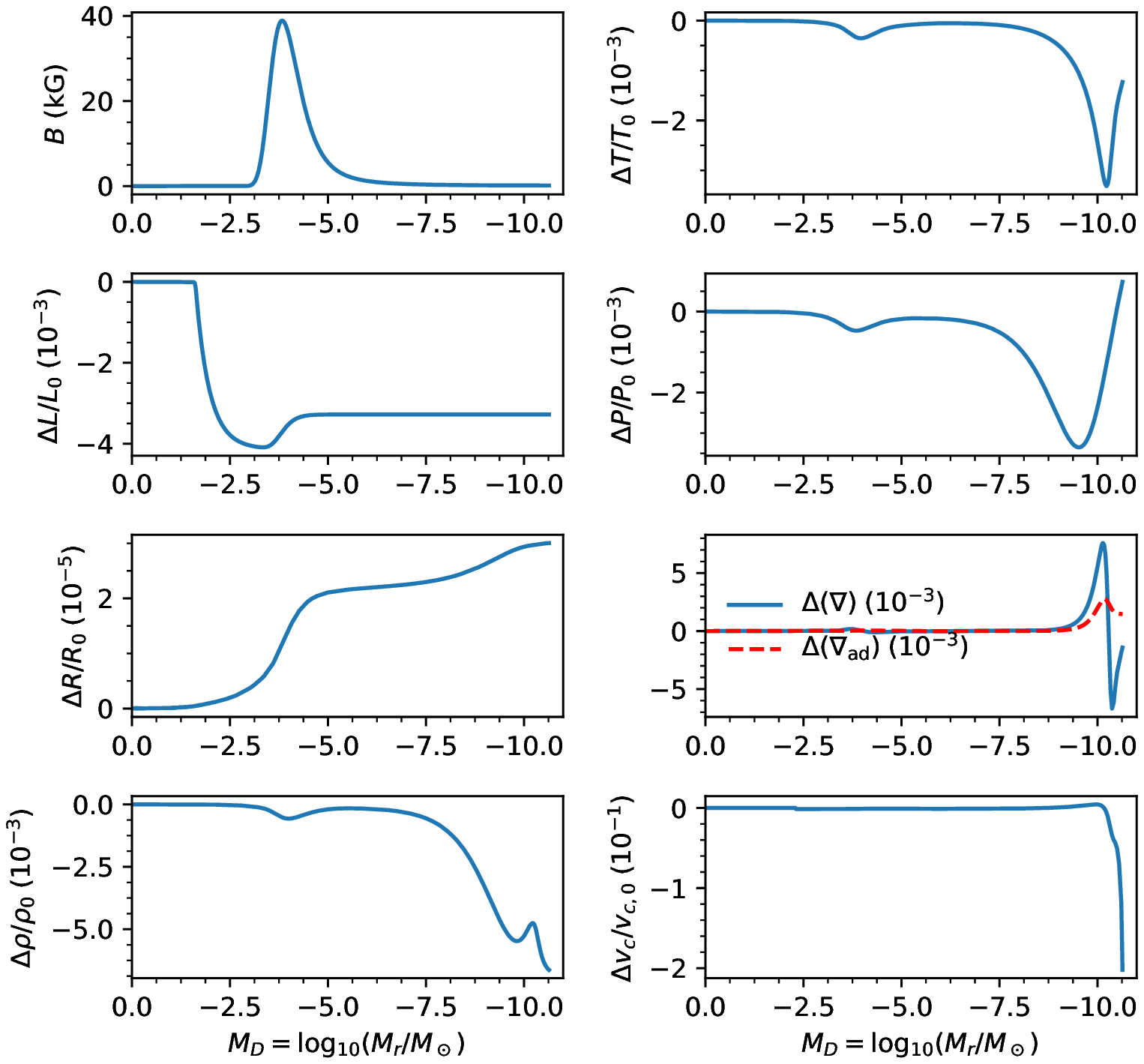}
\caption{Perturbations of the interior structure for the G4 model ($r_0=0.96$, $\sigma_0=0.012$, $B_0=38.9$ kG). 
The profile of the magnetic field is shown in the upper left panel; the other panels show the relative perturbations of the main structure variables. 
Note the inverted $x$ axis plotted in terms of the mass depth variable, $M_D$ (center: $M_D=0$, left; surface: $M_D\rightarrow -\infty$, right).}
\label{interior_gaussian}
\end{center}
\end{figure*}

\begin{figure*}
\begin{center}
\includegraphics[width=0.9\textwidth]{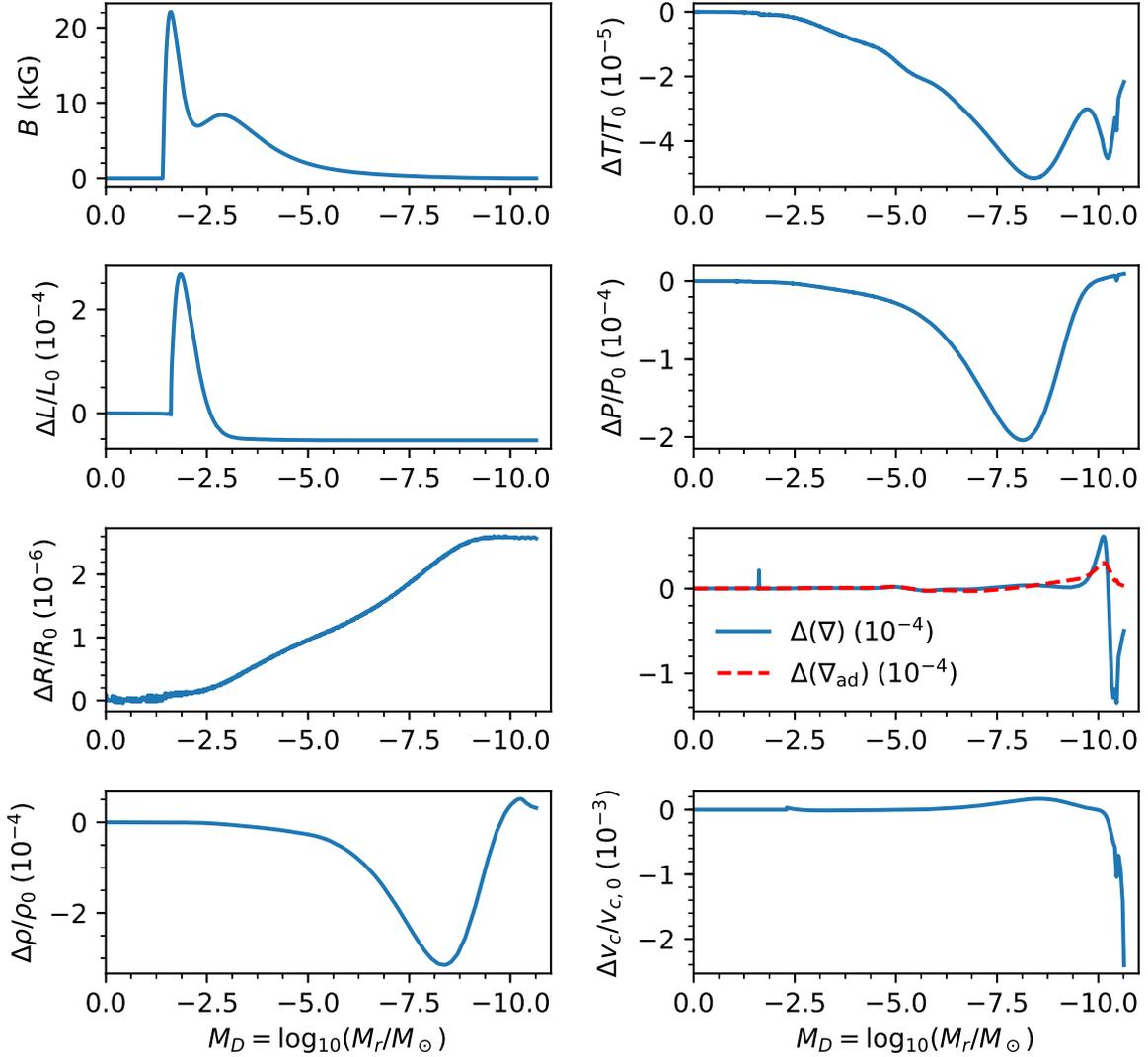}
\caption{Same as Figure \ref{interior_gaussian}, but for the DT model ($B_0=22.1$ kG): the magnetic field profile is shown in the upper left panel; in other panels, the relative perturbations of the main structure variables are plotted.}
\label{interior_dynamo}
\end{center}
\end{figure*}

\begin{table}
\begin{center}
\caption{Reference Standard Solar Model.}
\begin{tabular}{ccc}
\hline
\hline
Parameter & Adopted$^\dag$ & Model \\
\hline 
Age (Gyr) & $4.57$ & - \\
Mass (g) & $1.9891 \cdot 10^{33}$ & - \\  
Radius (cm) & $6.9598 \cdot 10^{10}$ & $\log(\frac{R}{R_\odot}) =  -1.20 \cdot 10^{-6}$ \\
Luminosity (erg/s) & $3.8418 \cdot 10^{33}$ & $\log(\frac{L}{L_\odot}) =-0.15 \cdot 10^{-6}$ \\ 
$R_{\rm bcz}$ ($R_\odot$) & $0.713$  &  $0.7149602$ \\
Surface $({Z/X})$ & $0.0230$ & $0.0229957$ \\
\hline
\end{tabular}
\label{ssm}
\end{center}
{\bf Notes.} 
$^\dag$ See, e.g., \cite{Basu_Antia:2008}.
\end{table}

\subsection{Non-magnetic reference model}

All our calculations are initialized from a SSM of age $t_\odot=4.57$ Gyr (by definition; the basic properties of our SSM are listed in Table~\ref{ssm}).

In magnetically perturbed runs, the perturbation is introduced at $t_0 \gtrsim t_\odot$, with either a step-like or a periodic  time dependence, equation \eqref{step} or \eqref{periodic} respectively. 
In a typical run, $t_0-t_\odot \approx 10$ yr, and we follow the evolution for $\approx 100$ yr (cf. Figure \ref{timedep}).
In order to fully resolve the dynamical effects of the magnetic field initial appearance and subsequent variations, we use time steps of $0.4$ yr.

Since the variations induced in both global parameters (i.e., surface radius, total luminosity, effective temperature) and local ones (run of pressure, density, temperature, etc., in the model interior) are expected to be quite small (between $10^{-6}$ and $10^{-3}$, as discussed in Section \ref{constraints}), it is important to define them with respect to a non-magnetic reference model: the precision of the stellar code is sufficient to distinguish such effects in a relative sense (see also the discussion in section 6 of \citealt{LS95}).
To this end, the evolution starting from our SSM and continuing with the magnetic field strength set to zero (all other parameters of the run, such as the time step, being the same) is used as reference.

\subsection{Effect of the perturbation on the interior structure}

Local perturbations to interior model variables are defined relative to the corresponding quantity in the non-magnetic model, at a fixed time step and as a function of the mass shell location within the model, e.g.,
\begin{equation*}
\Delta P(r) = P(r) - P_0(r),
\end{equation*}
where $P_0$ and $P$ refer to a non-magnetic and a perturbed model, respectively.
The perturbations are shown scaled to the local values, e.g., $\Delta P/P=[P(r)-P_0(r)]/P_0(r)$, except for the temperature gradients, $\nabla$ and $\nabla_{\rm ad}$.

In our formalism, magnetic fields affect the stellar structure in a variety of ways (cf. equations \ref{modP}--\ref{modE}); as a result, the perturbations are not always limited to the immediate neighborhood of the peak of the magnetic field distribution. 

Figure \ref{interior_gaussian} shows the local perturbations to the internal structure for the magnetic configuration of model G4.
The general features observed in the Figure are also representative of the other Gaussian models.

Since the magnetic field is sharply peaked, the perturbations of the main thermodynamic variables $P$, $T$, $\rho$, as well as of the convective velocity, are closely associated with the center of the magnetic layer.
The luminosity, on the other hand, is most affected by the changes near the bottom of the convection zone.
The radius perturbation is shaped like the integral of $f(r)$, i.e., an Error Function in this case, due to the fact that the expansion of each mass shell also affects all the layers above it, producing a cumulative effect towards the surface (see also figure 2 of \citealt{Sofia_ea:2005}). 
Finally, the perturbation of the temperature gradient roughly follows the derivative of $f(r)$, because the leading term of the perturbation in equation \eqref{modG} is proportional to $\nabla_\chi$ \citep[see also figure 1 of][]{Schatten_Sofia:1981}.

In contrast to the Gaussian models, the two dynamo profiles feature a deep-seated magnetic layer, peaking near the bottom of the convection zone, and a broad component with a gradual slope that extends towards the surface.
Since, in the normalization of $B_0$ according to equation \eqref{Emag}, models DB and DT have similar maximum field strength (cf. Table \ref{chiparm}), they differ mainly in the relative importance between the peaked, deep component and the broad, shallow one (see Figure \ref{chi_radial}).

For model DT, the interior perturbations are shown in Figure \ref{interior_dynamo}.
The effect of the moderate gradient of the magnetic field in the outer layers clearly dominates over that of the deep peak; this is especially evident in the case of the radius perturbation. 

The results for model DB are qualitatively similar, but smaller in magnitude, because they mostly arise from a magnetic field located at deeper layers, but of similar strength of that of model DT.

\subsection{Time scales of the reaction to the magnetic perturbation}
\label{timevol}

\subsubsection{Short term response}

At a given evolutionary time step, the changes to the total luminosity and the surface radius induced by the magnetic perturbation are:
\begin{equation}
\label{def_delta}
\delta {\cal L}(t) = \frac{{\cal L}(t) - {\cal L}_0(t)}{{\cal L}_0(t)},
\ \ \ \ 
\delta {\cal R}(t) = \frac{{\cal R}(t) - {\cal R}_0(t)}{{\cal R}_0(t)}; 
\end{equation}
where $\cal L$ and $\cal R$ denote surface luminosity and radius, respectively; the subscript $0$ is applied to the parameters of the non-magnetic model.

It is also useful to introduce $\cal W$, the ratio of the absolute value of the variations:
\begin{equation}
{\cal W} = |\delta {\cal R}/\delta {\cal L}|.
\label{def_w}
\end{equation}
This quantity is a measure of the importance of the radius perturbation relative to the luminosity perturbation; in other words, it measures whether the radius and the luminosity are equally affected by the presence of the magnetic fields.

The stellar structure equations are a set of highly coupled, non-linear equations, and their reaction to a perturbation occurs through a hierarchy of timescales \citep[see, e.g., chapter $25$ of][]{KWW12}. 
A perturbation of the hydrostatic equilibrium is restored within a dynamical timescale, which in the Sun is $\lesssim 1$ hour. 
The thermal energy transport by convection comes into equilibrium with the background stratification within a turnover timescale, $\approx 1$ month. 
A much longer timescale, of the order of $10^5$ yr, is required for complete thermal relaxation of the convection zone.
This composite time response, which results from the coupling between the variables in the stellar structure equations, is inherently dealt with by a standard stellar evolution code, such as YREC \citep[see the discussions in][]{Gough:1981,Gough:2002}. 

\begin{figure}
\begin{center}
\includegraphics[width=0.49\textwidth]{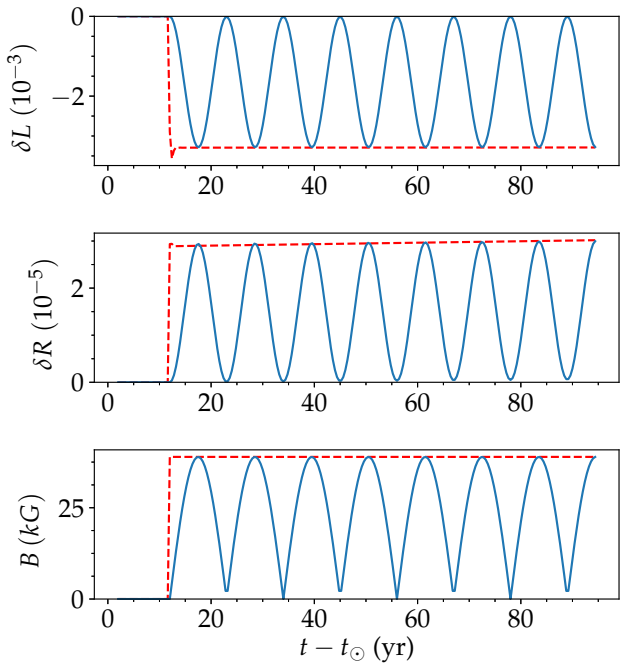}
\caption{Luminosity and radius variations vs. time for model G4 ($r_0 = 0.96$, $\sigma_0 = 0.012$, $B_0 = 38.9$ kG; $t_0=12$ yr, the time step is $0.4$ yr). Solid blue line: periodic time dependence; red dashed line: step-like time dependence.}
\label{timedep}
\end{center}
\end{figure}

The fastest components of the response, acting on $\approx$ hours to months timescales, cannot of course be fully resolved by the stellar code, and will therefore appear as instantaneous readjustments in our models (cf. Figure \ref{timedep} at $t \gtrsim t_0=12$ yr).
Nevertheless, the $0.4$ yr time step is sufficiently refined to resolve the $11$ yr period of the cycles in $B_0^2$.
Our main goal in this section is to assess whether the amplitude of the perturbations in $\delta {\cal L}$ and $\delta {\cal R}$ differ significantly between the step-like and the periodic time-dependences.

The response on short-term timescales ($\approx 10$s of years) is illustrated in Figure \ref{timedep}.
The structure reacts very quickly to the inception of the magnetic perturbation; this initial readjustment, as expected, is not fully resolved.
Indeed, in the case of a step-like perturbation, the equilibrium is restored within a time step.
In the periodic run, the radius and the luminosity follow the magnetic field oscillation essentially without phase lag.
The amplitude of $\delta {\cal L}$ and $\delta {\cal R}$ are essentially the same in the two cases. 

This result is not surprising: the very long response timescale of the convection zone as a whole ($\approx 10^5$ yr), which is a consequence of its very high thermal capacity, does not prevent the perturbation from being felt on the two faster timescales first, through the coupling with the other stellar structure equations (e.g., hydrostatic equilibrium).

\begin{figure}
\begin{center}
\includegraphics[width=0.5\textwidth]{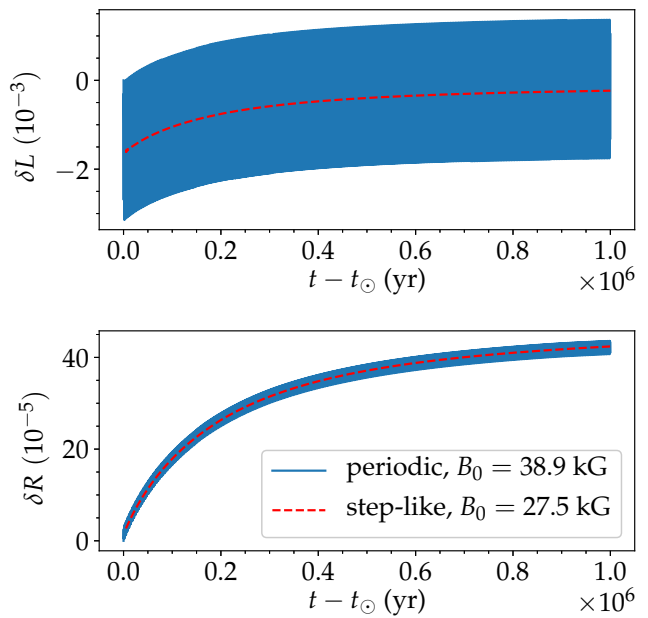}
\caption{Long-term luminosity and radius variations vs. time for model G4 ($r_0 = 0.96$, $\sigma_0 = 0.012$, $t_0=12$ yr, the time step is $0.4$ yr). Solid blue line: $B_0=38.9$ kG, periodic time dependence; the single oscillations with a period of $11$ yr, are not resolved at the scale of the plot. Red dashed line: step-like time dependence, $B_0=27.5$ kG (see text for discussion of the choice of $B_0$).}
\label{longterm}
\end{center}
\end{figure}

\subsubsection{Long term response}

To investigate the approach to the thermally relaxed state, we have constructed a longer evolutionary sequence, following the evolution for $10^6$ yr after $t_0$. 
This is shown in Figure \ref{longterm}, again for model G4 to facilitate the comparison with Figure \ref{timedep}.

The fast response to the perturbation produces oscillations of amplitude $\approx 3 \cdot 10^{-3}$ and $\approx 3 \cdot 10^{-5}$ in luminosity and radius, respectively (of course, neither the initial response nor the single oscillations are clearly distinguishable at the time resolution of Figure \ref{longterm}). 
A much slower readjustment of the structure follows, with an e-folding time of the order of $10^5$ yr, traced by the evolution of the average luminosity and radius, over which the oscillations are superposed. 

It is important to note that, while the average luminosity and radius slowly evolve, reaching an equilibrium value over the thermal timescale, the amplitude of the oscillations of either variable does not change much between the beginning and the end of the long-term run. 
A correct estimate of $\delta{\cal L}$ and $\delta{\cal R}$ can therefore be obtained without following the entire relaxation phase in detail. 

Asymptotically, the average luminosity and radius readjust to values that are almost unchanged, and slightly larger, respectively, in comparison to those of the unperturbed state.
Both results have a simple interpretation.
Since the magnetic perturbation does not directly affect the total energy output, the average luminosity variation tends to zero after complete thermal relaxation has occurred. 
The increase of the average radius, on the other hand, is a response to the extra pressure contribution, as well as to the additional obstacle to the convective energy flow provided by the magnetic fields.

In the same Figure, the periodic run for model G4 is compared with a step-like time dependence run for the same configuration, but with a value of $B_0$ reduced by a factor of $1/\sqrt{2}$.
The average luminosity and radius in the periodic run essentially coincide with their counterparts in the step-like run at reduced $B_0$.
This is a consequence of the magnetic field oscillations in the periodic run being averaged over the much longer thermal timescale ($11$ yr $\ll 10^5$ yr). 
Indeed, since the radius variation is proportional to $B^2$, the structure reacts to the average perturbation strength, $B_0^2/2$.

\subsection{Magnitude and sign of $\delta{\cal L}$ and $\delta {\cal R}$}

The global parameter variations obtained for the magnetic configurations presented in Section \ref{method} are summarized in Table \ref{results}.
From the results in the Table, we can draw the following general conclusions.

In all the models considered, the luminosity is found to decrease, and the radius to increase, in the presence of the magnetic perturbation (i.e., $\delta {\cal L}<0$ and $\delta {\cal R}>0$).
As a consequence, in the runs implementing the periodic time dependence of the magnetic field, equation \eqref{periodic}, the luminosity and radius variations are in anti-phase and in phase, respectively, with the field itself (cf. Figure \ref{timedep}).

\begin{table*}
\caption{Global effects of the magnetic perturbation. For the luminosity and radius variations, $\delta {\cal L}$, $\delta {\cal R}$, and $\cal W$ are defined in equations \eqref{def_delta} and \eqref{def_w}. For the effective temperature variations, $\Delta T_{\rm eff}=T_{\rm eff}-T_{\rm eff,0}$ (not normalized).}
\begin{center}
\begin{tabular}{ccccccc}
\hline
\hline
Model & $B_0$ (kG) & $\delta {\cal L}$ & $\delta {\cal R}$ & $\cal W$ & $\Delta T_{\rm eff} (K)$ \\
\hline
G1 & $19.4$ & $-7.5 \cdot 10^{-4}$ & $1.5 \cdot 10^{-6}$ & $2.0\cdot 10^{-3}$ & $-1.2$ \\
G2 & $20.8$ & $-8.7 \cdot 10^{-4}$ & $2.0 \cdot 10^{-6}$ & $2.3\cdot 10^{-3}$ & $-1.3$ \\
G3 & $26.2$ & $-1.4 \cdot 10^{-3}$ & $4.7 \cdot 10^{-6}$ & $3.3\cdot 10^{-3}$ & $-2.1$ \\
G4 & $38.9$ & $-3.3 \cdot 10^{-3}$ & $3.0 \cdot 10^{-5}$ & $9.1\cdot 10^{-3}$ & $-4.9$ \\
\hline
DB & $19.5$ & $-3.6 \cdot 10^{-4}$ & $2.2 \cdot 10^{-6}$ & $6.1\cdot 10^{-3}$ & $-0.53$ 
\\
DT & $22.1$ & $-5.2 \cdot 10^{-5}$ & $2.6 \cdot 10^{-6}$ & $5.0\cdot 10^{-2}$ & $-0.13$  \\ 
\hline
\end{tabular}
\end{center}
\label{results}
\end{table*}

We also find positive $\delta {\cal L}$ (and, more rarely, negative $\delta {\cal R}$) in some exploratory calculations run with very shallow-seated magnetic configurations ($r_0 > 0.990$). 
In these cases, however, the magnetic effect is very strongly dominated by the response of the near-surface layers alone, which are not well-resolved spatially in our models.  
We therefore regard these results as dubious and do not discuss them further.  

The amplitude of the luminosity variations is comparable, or even slightly exceeds, its observational upper limit of $10^{-3}$ (see Section \ref{constraints}).
For the radius, on the other hand, we always find changes well below our adopted upper limit ($2\cdot 10^{-4}$).

In general, the absolute magnitudes of the variations, $|\delta {\cal L}|$ and $|\delta {\cal R}|$, depend both on the strength and on the radial profile of the magnetic fields.  
Locally, the importance of the perturbation is controlled by the plasma beta parameter, $P_{\rm mag}/P$ (which in our models is always $\lesssim 10^{-2}$). 
A stronger field (i.e., larger $B_0$) is therefore required to produce variations of the same amplitudes for shallow-field configurations in comparison with deep-seated ones.

The effective temperature variations are also listed in Table \ref{results}.
For all the magnetic configurations discussed here, the effective temperature is decreased in the magnetically perturbed model; this effect is of the order of a few degrees K at most.

\begin{figure*}
\begin{center}
\includegraphics[width=0.49\textwidth]{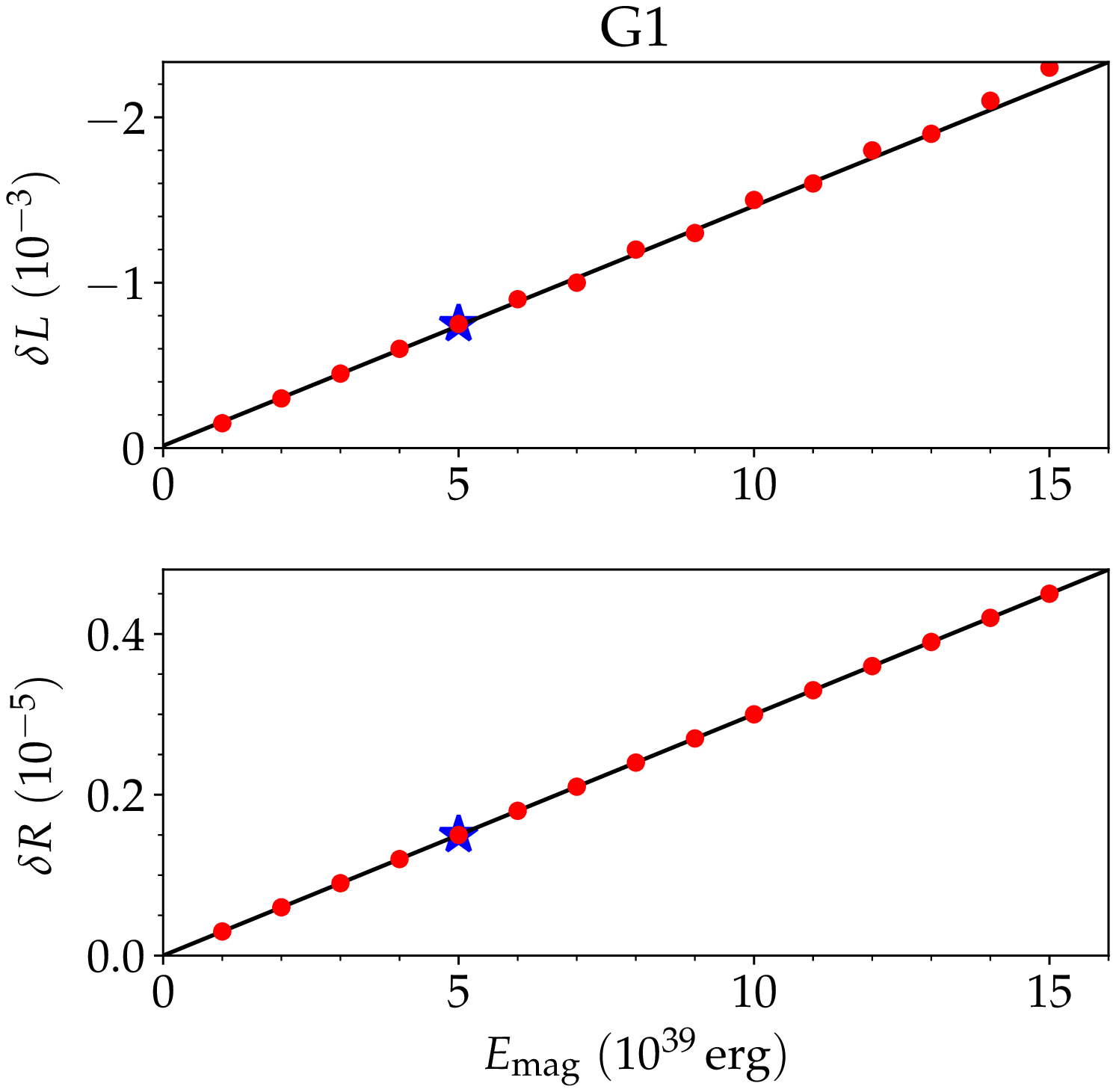}
\includegraphics[width=0.49\textwidth]{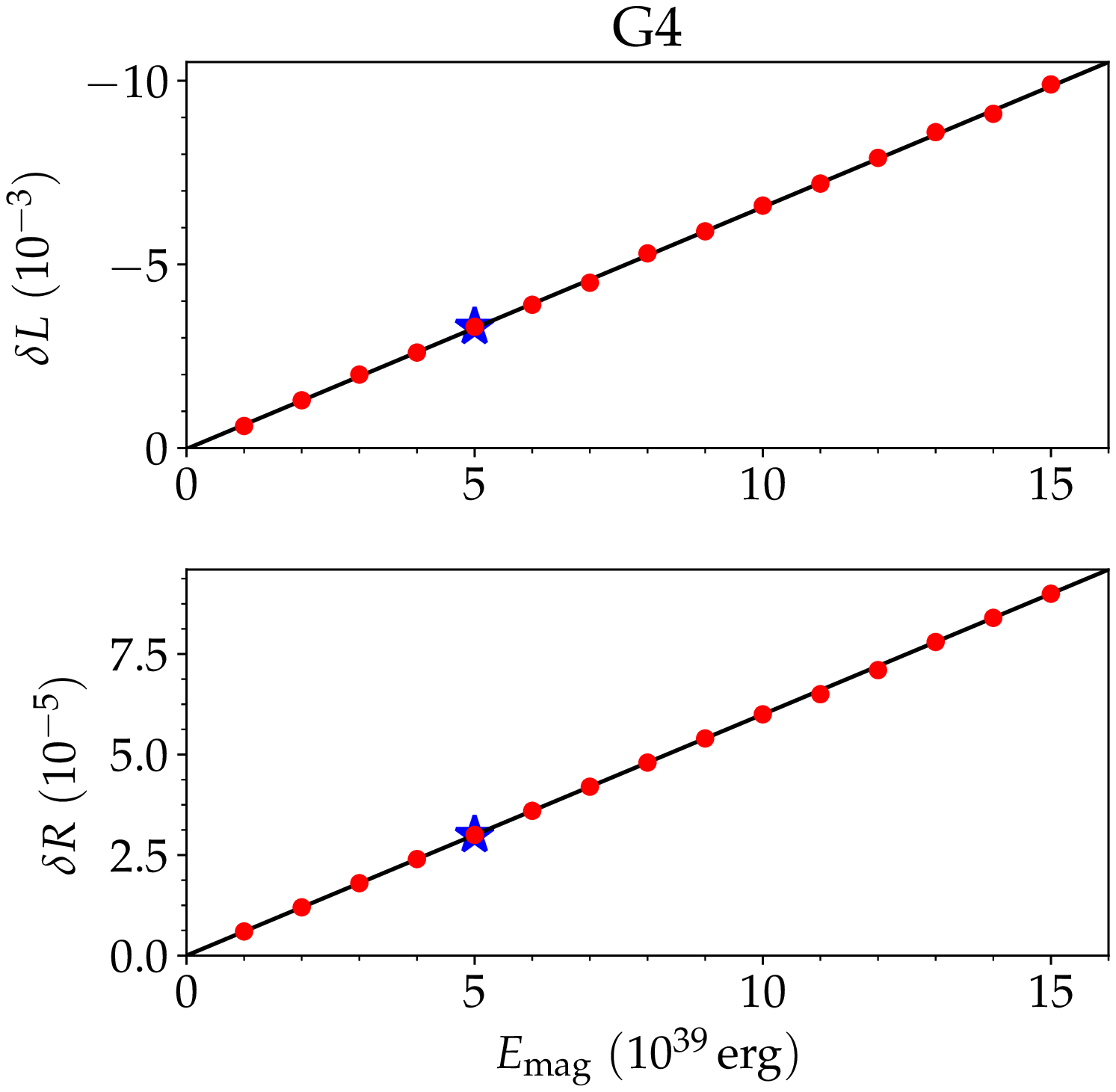}
\caption{Scaling of radius and luminosity variations vs. total magnetic energy for models G1 ($r_0 = 0.70$, $\sigma_0 = 0.090$, $B_0 = 19.4$ kG; left panels) and G4 ($r_0 = 0.96$, $\sigma_0 = 0.012$, $B_0 = 38.9$ kG; right panels); the black line represents a linear fit; the blue star marks the adopted case $E_{\rm mag}=5\cdot 10^{39}$ erg.}
\label{scaling}
\end{center}
\end{figure*}

Figure \ref{scaling} illustrates the dependence of $\delta {\cal L}$ and $\delta {\cal R}$ on $E_{\rm mag}$ for models G1 and G4.
This dependence is remarkably close to linear within the range of $E_{\rm mag}$ shown, as would be expected in the ideal perturbative regime.

An interesting consequence of this simple scaling is that ${\cal W}$ is independent of $E_{\rm mag}$: for example, the models shown in Figure \ref{scaling} all have $\langle {\cal W} \rangle=(2.0\pm0.04)\cdot 10^{-3}$ (G1) and $\langle {\cal W} \rangle=(9.2\pm0.2)\cdot 10^{-3}$ (G4; the standard deviations have been taken as an estimate of the uncertainties).
In other words, the relative importance of the luminosity and radius perturbations, expressed by the ratio $\cal W$, is entirely determined by the radial profile of the magnetic field alone, and does not depend on its peak strength $B_0$.

For both the Gaussian and the dynamo configurations, the value of $\cal W$ increases moving from deep-seated to shallow magnetic perturbations.
 This result could be used to constrain the location of the magnetic fields in the Sun if a reliable estimate of the $\cal W$ parameter was available. 
 
\section{Discussion}
\label{discussion}

The main goal of this work was to investigate the impact of magnetic fields in the interior of the Sun on its surface radius and total luminosity. 
In particular, we aimed to model the intrinsic effects associated with changes of the interior structure, and contrast the results with the well studied effect of the irradiance variations induced by surface magnetic phenomena \citep[e.g.,][]{Froehlich_Lean:2004,Froehlich:2013,Solanki_ea:2013}.

As discussed in the Introduction, constructing solar and stellar structure models with magnetic fields is a very complex task, and we have been forced to adopt a simplified approach.
In particular, although with some improvements tailored to the problem at hand, our calculation is done within the framework of a one-dimensional stellar evolution code (to be contrasted with, e.g., direct numerical simulations), and thus inherits all its intrinsic limitations. 
For example, we cannot follow the details of the energy exchanges between the magnetic fields, the plasma motions, and the thermal energy reservoir. 
However, this is a much more ambitious undertaking than our present goal, for which the approach adopted in this work is adequate.
Some of our simplifying assumptions are critically reviewed and further discussed below.

In our models, the most important perturbing effects arise from the magnetic contribution to the hydrostatic equilibrium, equations \eqref{modP} and \eqref{modD}, and from the inhibition of the energy transport via convection (equation \ref{modG}). 
Since we considered variable magnetic fields, their creation (dissipation) during a cycle acts as an energy sink (source) in the interior, and is thus included for consistency.
This effect is modeled as an extra term in the energy conservation equation, that we assume to be proportional to the time derivative of the total magnetic energy, and to be uniformly distributed over the entire convection zone (see equation \ref{modE}).
The treatment of the magnetic energy term as a perturbation is justified, because we only consider magnetic configurations with a total magnetic energy $E_{\rm mag} = 5\cdot 10^{39}$ erg.
This is much smaller than the total thermal energy content of the solar convection zone (which is of the order of $10^{46}$ erg, see, e.g., \citealt{Spruit:1982}).

Our treatment of the inhibition of the convective heat transport by magnetic fields implicitly assumes a ``frozen-in magnetic flux" condition. 
As a result, the energy transport effect resulting from magnetic fields being created in the deep interior and destroyed near the surface is not included.
This is, however, the main mechanism powering the surface magnetic activity and the surface-induced variations, which, as explained above, is an additional effect, besides the one that is the subject of our investigation.

The most severe limitation of our current modeling is the restriction to a one dimensional calculation.
Inherently 2D effects cannot be modeled within this approach: for example, if the magnetic field is confined to a toroidal structure of limited extension in a meridional plane, its impact on the heat transport by convection will not equally affect all latitudes, as in our 1D treatment.
Moreover, we cannot exploit in full the information provided by the two-dimensional dynamo models, and we have to assume that the latitudinal average of $B$ can capture, at least to the leading order, the essential features of the actual 2D configuration.
Ultimately, a fully 2D modeling, also including rotation and/or turbulence, will be required to incorporate a realistic magnetic field configuration in stellar models.

\section{Conclusions}
\label{conclusions}

We have constructed solar models that include the effects of the variable magnetic fields in its interior associated with the magnetic dynamo cycle.
We have studied several magnetic field configurations, formulated on the basis of the available observational constraints, as well as on the output of a mean field dynamo code.
The response to both step-like and periodic time dependences of the magnetic fields have been investigated.

In general, the magnetic models have larger radii and fainter luminosities with respect to their non-magnetic reference counterpart.
In runs implementing a magnetic perturbation whose strength varies periodically, the luminosity and radius variations are in anti-phase and in phase, respectively, with the magnetic field.

The amplitude of the luminosity perturbations is comparable to its observational upper limit ($10^{-3}$) in almost all the configurations considered, while the radius variations are more modest, $\lesssim 10^{-5}$, depending on the depth of the main magnetic layer.

The sensitivity of the results to the radial profile of the magnetic perturbation, and thus, in turn, to the details of the magnetic field amplification in the convection zone, suggests a novel potential avenue to constrain dynamo models.

Our results show that intrinsic luminosity variations, induced by the interior magnetic fields, can be of a similar order of magnitude as the observed TSI variation, which are currently successfully explained by models taking into account the effect of surface phenomena alone (i.e., sunspots, faculae, and magnetic network).
Since radius variations are associated with intrinsic structural variability only, future observations of the solar radius variability over a cycle, which are currently controversial, might help to disentangle these two effects and assess their relative importance.

\section*{Acknowledgements}

We thank an anonymous referee, whose thoughtful comments have helped us to improve this paper.

FS gratefully acknowledges support from the Yale Institute for Biospheric Studies (YIBS) from 2010 to 2012, when the foundations of this work were laid, and from the Karl Schwarzschild Postdoctoral Fellowship at AIP (2012 to 2017). 
SS acknowledges the generous support of the G. Unger Vetlesen Foundation and of the Brinson Foundation over a period sufficiently long to allow us to carry to successful conclusion both the theoretical and the experimental work required to define and explore a mechanism of solar variability that can affect climate.

\bibliographystyle{mn2e}


\onecolumn
\appendix

%
%
%
%

\section{Changes to the thermodynamic derivatives in the presence of magnetic fields}

\label{app_derivatives}

\subsection{Original \citet{LS95} treatment}

The original \citet{LS95} treatment relied on the following idealized equation of state (EOS) and energy density equation:
\begin{equation}
\label{ls95}
P_T = \frac{\cal R}{\mu} \rho T + \frac{1}{3}aT^4 + (\gamma-1) \rho \chi;
\ \ \
u_T = \frac{3}{2} \frac{\cal R}{\mu} T + \frac{aT^4}{\rho} + \chi,
\end{equation}
where $P_T$ and $u_T$ are the total pressure and energy density, i.e., including the contributions from the gas, radiation, and magnetic fields. 
The latter is expressed in terms of the magnetic energy density per unit mass $\chi$, while the factor $(\gamma-1)$ takes phenomenologically into account the effects of the magnetic tension: 
\begin{equation*}
P_\chi = (\gamma-1) \rho \chi = (\gamma-1) \frac{B^2}{8\pi}.
\end{equation*}
In the following, we set $\gamma=2$, i.e. we neglect the magnetic tension.

In a stellar evolution code like YREC, the EOS routines are called to supply the density as a function of pressure, temperature, and chemical composition, along with their partial derivatives:
\begin{equation}
\label{eos}
 \dfrac{d\rho}{\rho}  =  \alpha \dfrac{dP}{P} - \delta \dfrac{dT}{T}, \ \mathrm{with}
\ \
 \alpha \equiv \left( \dfrac{\partial \ln \rho}{\partial \ln P} \right)_T;
\ \ 
 \delta \equiv - \left( \dfrac{\partial \ln \rho}{\partial \ln T} \right)_P . 
\end{equation}
In the presence of the magnetic perturbation, the new state variable $\chi$ explicitly enters in the EOS:
\begin{equation}
\label{eosmod}
\dfrac{d\rho}{\rho} = \tilde\alpha \dfrac{dP_T}{P_T} - \tilde\delta \dfrac{dT}{T} - \nu \dfrac{d\chi}{\chi};
\ \mathrm{with} \ \
\tilde \alpha \equiv \left( \dfrac{\partial \ln \rho}{\partial \ln P_T} \right)_{(T,\chi)};
\ \ 
\tilde \delta \equiv - \left( \dfrac{\partial \ln \rho}{\partial \ln T} \right)_{(P_T,\chi)};
\ \ 
\nu \equiv - \left( \dfrac{\partial \ln \rho}{\partial \ln \chi} \right)_{(P_T,T)} .
\end{equation}
From the definition of $P_T$ in equation \eqref{ls95}:
\begin{equation*}
\rho = \frac{P_T - \dfrac{1}{3}aT^4}{\dfrac{\cal R}{\mu}T + \chi},
\ \Rightarrow \ \
\tilde\alpha = \dfrac{P_T}{P_T - \dfrac{1}{3}aT^4};
\ \  
\tilde\delta = \dfrac{\dfrac{4}{3}aT^4 + \dfrac{\cal R}{\mu}\rho T}{P_T - \dfrac{1}{3}aT^4};
 \ \ 
 \nu = \dfrac{\rho\chi}{P_T - \dfrac{1}{3}aT^4}.
\end{equation*}

\subsection{This work}

The formulation used in this work retains the full generality of the EOS options normally used in YREC (e.g. the OPAL 2005 EOS, \citealt{Iglesias_Rogers:1996}), without resorting to the ideal gas EOS in equation \eqref{ls95}. 
The difference is minor in the case of the Sun and for deep perturbations, but it could be more relevant for stars of lower mass and/or for perturbations localized near an ionization zone.

We start by writing the total pressure as $P_T = P + P_\chi$ (recall that $P_\chi = \rho \chi$ and thus $\ln P_\chi = \ln \rho + \ln \chi$):
\begin{equation*}
 \dfrac{dP_T}{P_T} = \dfrac{dP + dP_\chi}{P_T} = \dfrac{P}{P_T} \dfrac{dP}{P} + \dfrac{P_\chi}{P_T} \dfrac{dP_\chi}{P_\chi}  =  \dfrac{P}{P_T} d\ln P + \dfrac{P_\chi}{P_T} \left( d \ln \rho + d\ln \chi \right)
\end{equation*}
thus:
\begin{equation*}
d\ln P = \dfrac{P_T}{P} d\ln P_T  - \dfrac{P_\chi}{P} (d\ln \rho + d\ln \chi);
\end{equation*}
inserting this result in \eqref{eos}, we have:
\begin{eqnarray*}
d\ln\rho = \alpha \left[   \dfrac{P_T}{P} d\ln P_T  - \dfrac{P_\chi}{P} (d\ln \rho + d\ln \chi) \right] - \delta \dfrac{dT}{T} 
\end{eqnarray*}
hence:
\begin{equation*}
\left( 1 + \alpha \dfrac{P_\chi}{P} \right) d\ln \rho = \alpha \dfrac{P_T}{P} d\ln P_T  - \delta d\ln T - \alpha \dfrac{P_\chi}{P} d\ln \chi, 
\end{equation*}
or
\begin{equation*}
d\ln\rho =  \frac{ \alpha \frac{ P_T}{P}}{1 + \alpha \frac{P_\chi}{P}} d\ln P_T -  \frac{\delta}{1 + \alpha \frac{P_\chi}{P}} d\ln T - \frac{\alpha \frac{P_\chi}{P}}{1 + \alpha \frac{P_\chi}{P}} d\ln \chi,
\end{equation*}
and, comparing with \eqref{eosmod}, we have:
\begin{equation}
\label{myalpdel}
\frac{\tilde \alpha}{\alpha} = \frac{P_T}{P + \alpha P_\chi}; 
\ \ \ \  
\frac{\tilde\delta}{\delta} = \frac{P}{P+\alpha P_\chi}; 
\ \ \ \ 
\frac{\nu}{\alpha} = \frac{P_\chi}{P+\alpha P_\chi}.
\end{equation}
These equations give the modified thermodynamic derivatives $\tilde\alpha$, $\tilde\delta$, and $\nu$ as a function of known quantities, namely the unperturbed $\alpha$ and $\delta$ and the gas, magnetic, and total pressure.
The relation $\frac{\nu}{\tilde\alpha} = \frac{P_\chi}{P_T}$ will also be useful in the following.
%

\end{document}